# Anyonic interference and braiding phase in a Mach-Zehnder Interferometer

Hemanta Kumar Kundu, Sourav Biswas, Nissim Ofek[1], Vladimir Umansky, Moty Heiblum[*]

*Braun Center for Submicron Research, Department of Condensed Matter Physics,*

*Weizmann Institute of Science, Rehovot, Israel*

[1]*Quantum Machines, 126 Yigal Alon St., Tel Aviv, Israel*

[*]E-mail: moty.heiblum@weizmann.ac.il

**The fractional quantum Hall states have long been predicted to be a testing ground of fractional (anyonic) exchange statistics. These topological states harbor quasiparticles with fractional charges of both abelian and non-abelian characters. The quasiparticles' charge is commonly determined by shot noise measurements (*1, 2*), and states' statistics can be revealed by appropriately interfering the quasiparticles. While the multipath Fabry-Perot electronic interferometer (FPI) is easier to fabricate, it is often plagued by Coulomb interactions (*3*), its area breathes with the magnetic field (*4*), and its bulk's charges tend to fluctuate (*5*). Recent FPI experiments employing adequate screening allowed an observation of Aharonov-Bohm (AB) interference at bulk filling $\nu$=1/3 (*6*). In the current work, we chose to employ an interaction-free, two-path, Mach-Zehnder interferometer (MZI), tuned to bulk filling $\nu$=2/5. Interfering the outer $\nu$=1/3 mode (with the inner $\nu$=1/15 mode screening out the bulk), we observed a 'dressed AB' periodicity, with a combined 'bare AB' flux periodicity of three flux-quanta ($3\phi_0$) and the 'braiding phase' $2\pi/3$. This unique interference resulted with an AB periodicity of a single flux-quantum. Moreover, the visibility of the interference, $v_{e/3}$, deviated markedly from that of the electronic one $v_e$, agreeing with the theoretically expected visibility, $v_{e/3} \sim v_e^3$. With the two non-equivalent drains of the MZI, the fractional visibility peaked away from the ubiquitous transmission-half of the MZI. We provide simple theoretical arguments that support our results. The MZI proves to be a powerful tool that can be used to probe further the statistics of more complex anyonic quasiparticles.**

Quantum Hall states were the earliest protagonists of topological phases of matter. While the bulk is insulating, the current is carried by chiral edge modes with a universal edge conductance of $\nu e^2/h$, where $e$ is the electron charge, $h$ the Planck's constant, and $\nu$ the filling factor (integer or fraction) (*7-9*). In the fractional quantum Hall effect (FQHE), the excitations are quasiparticles (QPs) that carry fractional charges (*10-13*). The QPs are neither bosons nor fermions; they are classified as *anyons* (*14, 15*). Upon exchanging two identical anyons, the phase of their joint wavefunction changes by a fraction of π, whereas it is π (2π) for fermions (bosons) (*16-18*).

The straightforward method to study the anyonic statistics of the QPs is to interfere the edge modes around localized QPs, thus performing a braiding operation. The two well-studied interferometers are the electronic Fabry-Perot Interferometer (FPI) (*3, 4, 6, 10, 19-24*) and the electronic Mach-Zehnder Interferometer (MZI) (*25-31*). While the bare FPI (a large version of a 'quantum dot') possesses finite charging energy (for the addition of QPs) (*4, 32, 33*), which tends to affect the interference, sufficient screening already enabled observation of fractional Aharonov-Bohm (AB) interference (*6*). On the contrary, the MZI is free of charging effects since one of its (grounded) drains is located in its interior, thus adding or removing particles at will. Yet, thus far, AB interference was observed only in the integer QHE regime (*25-27, 34*). The apparent lack of anyonic interference was attributed to the relatively larger interferometer size, the poor quality of interior (small) drain contact, and the presence of non-topological neutral mode (*35-41*).

Here, we describe the first observation of high visibility interference of the outer $\nu=1/3$ edge mode in bulk filling factor $\nu=2/5$, employing an optimized MZI. As we detail below, the MZI is unique because the observed AB interference is naturally 'dressed' by an additional anyonic braiding. Below, we describe the interferometer structure, the experimental results, and the theoretical analysis.

The Mach-Zehnder interferometer was formed by two closely placed quantum point contacts (QPCs), acting as 'beam splitters' and two ohmic drains: D2 - a small, grounded, set on the inner periphery of the MZI; and D1 - set downstream from the MZI (see Figs. 1a & 1b). The incoming charged edge mode splits in QPC1 into two trajectories that rejoin in QPC2, enclosing a magnetic flux. Note that there is a π phase difference between the reflection and transmission amplitudes in each QPC. With each QPC's transmission (reflection) amplitude $t_i$ ($r_i$), the two transmission probabilities of the MZI in the integer regime are, $T_{D1} = |t_1 t_2 + r_1 r_2 e^{i2\pi\phi/\phi_0}|^2 = |t_1 t_2|^2 + |r_1 r_2|^2 + 2|t_1 t_2 r_1 r_2|\cos(2\pi\phi/\phi_0)$, and $T_{D2} = |t_1 r_2 + r_1 t_2 e^{i2\pi\phi/\phi_0}|^2 = |t_1 r_2|^2 + |r_1 t_2|^2 - 2|t_1 t_2 r_1 r_2|\cos(2\pi\phi/\phi_0)$, where $\phi/\phi_0$ is the number of flux-quanta threading the effective interferometer area (the area enclosed by the two trajectories), and $T_{D1} + T_{D2} = 1$. A 'modulation gate' (MG) tunes the threaded flux via changing the enclosed area. The electrons' visibility is $v_e = \frac{T_{\max} - T_{\min}}{T_{\max} + T_{\min}}$, where $T_{\max}$ ($T_{\min}$) is the maximum (minimum) transmission at each drain.

We studied two different-size MZIs. One with an 'effective areas' ~ 3.67μm² and a larger one with an area ~13.5μm² (Fig. 1a), and single path lengths ~1.9μm and ~5.1μm, respectively. The interferometers were fabricated in a high mobility two-dimensional electron gas embedded in a GaAs/AlGaAs heterostructure (grown by MBE in our center). We tested two different MBE growths: electron densities (0.92, 1.22)×10¹¹cm⁻², and 4.2K dark-mobility (4.1, 3.6)×10⁶cm²/V-s, respectively. Ohmic contacts and gates were formed standardly (see Supp. Sec. SM1), and measurements were conducted at electrons' base temperature 10–15mK. The conductance and shot noise were measured at ~900kHz, with an appropriate bandwidth in both cases. A homemade amplifier, cooled to 1.5K, cascaded by a room temperature amplifier, provided a total gain of ~5000. The measurement results are summarized in the **Table** below.

We started with the smaller MZI tuned to bulk filling factor, *ν*=3 and *ν*=2. The interference *pajama plot* (conductance in $V_{MG}$–*B* plane) of the outer edge mode in *ν*=2 is shown in Fig. 2a. The plot is characteristic of constant-area AB interference. The *B*-dependence flux periodicity is the flux-quantum, $\phi_0 = h/e$ - as expected (Figs. 2a & 2b and **Table**; see also Supp. Sec. SM2 & Fig. SM2). In both *ν*=3 & *ν*=2, the most inner edge modes were dephased. The general behavior of the large MZI was similar to that of the smaller one; only, with visibility reduced (see **Table** and Supp. Sec. SM8).

Charging negatively the modulation gate (MG) depletes charge from the edge, $\Delta q = C\Delta V_{MG} = n_e e \Delta A$, and consequently reduces the AB phase, $\Delta\theta = 2\pi \frac{B \Delta A}{\phi_0}$. A constant gate capacitance *C* in all filling factors leads to fluxes ratio, $\frac{\Delta\phi_2}{\Delta\phi_1} = \frac{B_2 \Delta V_{MG2}}{B_1 \Delta V_{MG1}}$ (see Supp. Sec. SM3). The expelled charges per interference period are $\Delta q(\nu=2)=2e$ and $\Delta q(\nu=3)=3e$; corresponding to the removal of a single flux quantum per period (see **Table**).

| device size 2D density | $\nu$ | $B$ (Tesla) | $\Delta B$ (Gauss) | $\Delta\phi$ ($\phi_0$) | $\Delta V_{MG}$ (mV) | $\Delta\|q\|$ ($e$) | $\frac{\Delta\phi_2}{\Delta\phi_1} = \frac{B_2 \Delta V_{MG2}}{B_1 \Delta V_{MG1}}$ | visibility (%) |
|---|---|---|---|---|---|---|---|---|
| 3.67μm² high density | 2 | 2.5 | 10.8 | 1 | 22.2 | 2 | 1 | 90.7 |
| | 3 | 1.675 | 10.96 | 1.01 | 32.9 | 2.96 | 1.01 | 62.4 |
| | 2/5 | 12.65 | 10.59 | 0.98 | 4.2 | 0.38 | 1.04 | 22.0 |
| 13.5μm² low density | 2 | 1.85 | 3.02 | 1 | 9.8 | 2 | 1 | 67.6 |
| | 3 | 1.245 | 3.02 | 1.0 | 14.7 | 3.0 | 0.99 | 13.9 |
| | 2/5 | 9.05 | 2.91 | 0.96 | 1.8 | 0.37 | 1.10 | 8.3 |

**Table**: Details of the interference at the integer and fractional QHE regimes. The normalization is with respect to the interference of the outer edge in $\nu=2$ with electron interference periodicity of $\phi_0$.

The highest visibility, measured at $T_{MZI-D1} \leq 0.5$ (distinguishing from the maximum interference amplitude found at $T_{MZI-D1}=0.5$), was ~91%. With increasing the magnetic field, the visibility gradually diminished, ultimately disappearing at $\nu=1$ (Supp. Sec. SM4). Such dependence was attributed to the emergence of non-topological neutral modes, resulting from spontaneous edge-reconstruction (*35, 36*).

Moving to the fractional regime, we studied the interference of the outer edge mode in bulk filling $\nu=2/5$. The state supports two downstream edge modes: an inner mode with conductance $e^2/15h$ and an outer mode with conductance $e^2/3h$. (see Supp. Sec. SM5 and Fig. SM4). The two QPCs were tuned to weakly partition the 1/3 mode and fully backscatter the 1/15 mode. The partitioned 1/3 mode carried shot noise. We measured Fano factor =1/3, corresponding to partitioned quasiparticles charge $e/3$ (Fig. 3a and Supp. Sec. SM6).

Counter to simplistic expectations, the observed periodicity of the *B*-dependence flux was of one flux-quantum (Fig. 2c and **Table**). The $V_{MG}$ periodicity corresponded to a depleted charge $\Delta q \cong 0.4e$ per period; corresponding to the removal of a single flux-quantum (**Table** and Supp. Sec. SM3). Similar data were obtained with the larger MZI (see **Table** and Supp. Sec. SM8). The 1/3 interference diminished exponentially with increasing temperature, with a characteristic temperature of ~23mK (Figs. 3b & 3c). Comparing the visibilities in the two MZIs, we estimate a dephasing length of 10.5µm for electrons and 3.3µm for *e*/3 QPs.

The present data of interfering *e*/3 quasiparticles (confirming the theory (*21, 42*)) proves that the 'anyonic MZI' behaves dramatically differently from the 'anyonic FPI'. In the FPI: the *B*-dependence flux periodicity is $3\phi_0$, whereas the $V_{MG}$ repels $\Delta q = e$ per-period; corresponding to three *e*/3 depleted from the interfering Landau level (LL) (*6*). In the MZI: the *B*-dependence flux periodicity is $\phi_0$, whereas $V_{MG}$ repels $\Delta q = \nu e$ per-period; corresponding to a single *e*/3 depleted from the interfering LL.

The observed flux periodicity of (an integer like) $\phi_0$ in the 'anyonic MZI' manifests the 'dressed AB' interference, which combines a 'bare AB' interference with an anyonic braiding. It will become more evident as we explore the visibility with the average transmission of the MZI. For electrons, say in the downstream drain D1, $T_{MZI-D1} = <T_{D1}> = |t_1 t_2|^2 + |r_1 r_2|^2$, and the visibility, $v_e = \frac{2\eta |t_1 t_2 r_1 r_2|}{T_{MZI-D1}}$, with $\eta$ a dephasing factor. The visibility of the electrons increases smoothly from zero at unity transmission to a maximum at transmission half, namely, $T_{MZI-D1} = T_{MZI-D2} = 0.5$. Below this transmission, it remains constant (Fig. 4f). However, in the fractional regime, the observed visibility at D1 ($v_{e/3}$) changes sharply with the MZI's average transmission, and it peaks at $T_{MZI-D1} > 0.5$ ($T_{MZI-D2} < 0.5$), with the two drains affecting markedly differently the AB interference (Fig. 4g). Before presenting the corresponding data in more detail, we provide a somewhat pictorial picture of the interfering *e*/3 QPs (*30, 31, 43-45*).

The 'picture' is based on the notion that the inner drain can absorb only electrons. When the AB phase directs a QP to the inner drain (D2), the QP charge is screened by holes in the drain; however, its attached flux remains uncompensated (occupying an 'area' of a flux quantum). For each of the next two quasiparticles, a braiding phase of $\frac{2\pi}{3}$ is added to the 'bare AB' anyonic phase. Hence, the evolving 'dressed AB' phase is $2\pi\left(\frac{\phi}{3\phi_0}+\frac{N-1}{3}\right)$, with $N^{th}$ interfering QPs entering the MZI and $n=(N-1)$ fluxes are stuck at the inner drain. The $N=4$ QP will be equivalent to the $N=1$ QP, as the phase will return to its initial value modulo 3. After three consecutive QPs accumulate at the drain, they leave to the ground as an electron. This mechanism has an extraordinary consequence on the visibility and periodicity of the interference pattern, as we describe now (more details in Supp. Sec. SM7).

The 'time' required for the $(n+1)^{th}$ quasiparticle to arrive at D2 is $t_n = 1/p_n$, where $p_n$ is its probability of arriving at D2. This probability is: $p_n = \bar{p}(1 + v_e \cos 2\pi(\frac{\phi}{3\phi_0}+\frac{n}{3}))$, with $\bar{p}$ the flux-independent probability determined by $t_i$'s of the QPCs (*30*),

$$T_{D2} \equiv 3\left[\sum_{n=0}^{2}\frac{1}{p_n}\right]^{-1} = 3\bar{p}\sum_{n=0}^{2}\left[\frac{1}{1+v_e \cos\left(\frac{2\pi\phi}{3\phi_0}+\frac{2\pi n}{3}\right)}\right]^{-1} = T_I\frac{4-3v_e^2}{4-v_e^2}\left[1+\frac{v_e^3}{4-3v_e^2}\cos\left(\frac{2\pi\phi}{\phi_0}\right)\right], \quad (1)$$

where $T_I = |t_1 r_2|^2 + |r_1 t_2|^2$. With current conservation, the transmission to D1,

$$T_{D1} = \left(1 - T_I\frac{4-3v_e^2}{4-v_e^2}\right) - T_I\frac{v_e^3}{4-v_e^2}\cos\left(\frac{2\pi\phi}{\phi_0}\right). \quad (2)$$

A few essential features of the anyonic interference are apparent: **1.** The flux periodicity is $\phi_0$, being consistent with Byers–Yang theorem (*46*); **2**. The two drains are not equivalent since the interfering quasiparticle encircles only D2; **3.** The oscillation amplitude is substantially lower than in the integer regime; **4.** The visibility at D1, $v_{e/3} = T_I v_e^3/[(4-v_e^2) - T_I(4-3v_e^2)]$, reaches a maximum value at average $T_{MZI-D1}>0.5$; hence, not the same as in D2, where $T_{MZI-D2}<0.5$. These values depend on $v_e$.

For example, for $v_e = 1$, the calculated fractional visibility, $v_{e/3}$, has a maximum at $T_{MZI-D1}=0.83$ and $T_{MZI-D2}=0.17$. Though the oscillation amplitude in both drains is the same, the visibility (accordingly to its definition) is: $v_{e/3}\sim20\%$ at D1 and $v_{e/3}=100\%$ at D2 (Fig. 4a). Moreover, with decreasing $v_e$, the disparity between D1 and D2 becomes less evident. Calculated illustrative examples of the interference patterns in both drains are shown in Fig. 4a.

We turn our attention to the experimentally observed visibility profiles in the small MZI. The interference pattern of the outer edge mode at filling $v=2$ is plotted in Fig. 4b. As many combinations of the two individual transmissions $|t_i|^2$'s of the two QPCs lead to the same average transmission of the MZI, we plot the visibility for many of these combinations (Fig. 4d). With a dephasing factor $\eta = 0.91$ in $v=2$, the calculated visibility plotted in Fig. 4f agrees in its general behavior with the measured one. The ponts following each line are for single $|t_1|^2$, while $|t_2|^2$ is varied from 0 to 1. This way for different $|t_1|^2$, configurations spanning all possiblities are studied. The AB interference of the larger MZI, having $\eta = 0.67$, is plotted in Supp Fig. SM6.

The interference pattern of the outer 1/3 mode in $v=2/5$ is plotted in Fig. 4c, with the visibility, $v_{e/3}$, plotted for many $|t_i|^2$'s as a function of $T_{\text{MZI-D1}}$ in Fig. 4e. The visibility peaks sharply with a maximum value $v_{e/3} \sim 22\%$ at $T_{\text{MZI-D1}} \sim 0.6$, decreasing rapidly on either side of the peak. Using Eq. 2 and $\eta = 0.91$, the calculated visibility follows the same behavior as the measured one (Fig. 4g). However, noteworthy the discrepancy between theory and experiment, where the theoretical visibility peaks at $T_{\text{MZI-D1}} \sim 0.76$ at $v_{e/3} \sim 15.6\%$, suggesting that other factors must be included in the analysis. The larger MZI behaved similarly, with peak visibility, $v_{e/3} \sim 8.3\%$ at $T_{\text{MZI-D1}} \sim 0.54$. See more details in the **Table** (see Supp. Sec. SM8).

Employing an electronic two-path Mach-Zehnder interferometer (MZI) in the fractional filling $v=2/5$, with the outer edge mode $v=1/3$ interfering, we demonstrated its sensitivity to braided quasiparticles tied to an additional accumulated Aharonov-Bohm phase. The observed 'dressed Aharonov-Bohm' interference of $e/3$ quasiparticles exhibited a markedly different fingerprint than the ubiquitous interference of electrons. With partitioned charges $e/3$, the two drains (inner and outer) were highly non-equivalent, thus leading to an uncharacteristic evolution of the interference visibility. Most importantly, the periodicity with magnetic field conspired to be a single flux-quantum; however, with area tuning (via a modulation-gate), each period corresponded to evacuation of a single $e/3$ change (one flux quanta) from interfering edge mode. This first step in studying an abelian anyonic interference in a two-path Mach-Zehnder interferometer, opens the field for more challenging interference experiments with more exotic (i.e., non-abelian) quantum Hall states.

**References:**

1. R. dePicciotto *et al.*, Direct observation of a fractional charge. *Nature* **389**, 162-164 (1997).
2. L. Saminadayar, D. C. Glattli, Y. Jin, B. Etienne, Observation of the e/3 fractionally charged Laughlin quasiparticle. *Physical Review Letters* **79**, 2526-2529 (1997).
3. N. Ofek *et al.*, Role of interactions in an electronic Fabry–Perot interferometer operating in the quantum Hall effect regime. *Proceedings of the National Academy of Sciences* **107**, 5276-5281 (2010).
4. I. Sivan *et al.*, Observation of interaction-induced modulations of a quantum Hall liquid's area. *Nature Communications* **7**, 12184 (2016).
5. B. Rosenow, S. H. Simon, Telegraph noise and the Fabry-Perot quantum Hall interferometer. *Physical Review B* **85**, 201302 (2012).
6. J. Nakamura *et al.*, Aharonov–Bohm interference of fractional quantum Hall edge modes. *Nature Physics* **15**, 563-569 (2019).
7. B. I. Halperin, Quantized Hall conductance, current-carrying edge states, and the existence of extended states in a two-dimensional disordered potential. *Physical Review B* **25**, 2185-2190 (1982).
8. X.-G. Wen, in *Quantum field theory of many-body systems: From the origin of sound to an origin of light and electrons.* (Oxford University Press on Demand 2004).
9. C. L. Kane, M. P. A. Fisher, in *Perspectives in Quantum Hall Effects: Novel Quantum Liquids in Low-Dimensional Semiconductor Structures,* S. Das Sarma, A. Pinczuk, Eds. (John Wiley, New York, 1996).
10. A. Stern, Anyons and the quantum Hall effect - A pedagogical review. *Annals of Physics* **323**, 204-249 (2008).
11. H. Z. Zheng, H. P. Wei, D. C. Tsui, G. Weimann, Gate-controlled transport in narrow $GaAs/Al_xGa_{1-x}As$ heterostructures. *Physical Review B* **34**, 5635 (1986).
12. M. Heiblum, D. E. Feldman, Edge probes of topological order. *International Journal of Modern Physics A* **35**, (2020).
13. A. M. Chang, Chiral Luttinger liquids at the fractional quantum Hall edge. *Reviews of Modern Physics* **75**, 1449-1505 (2003).
14. J. M. Leinaas, J. Myrheim, On the theory of identical particles. *Il Nuovo Cimento B (1971-1996)* **37**, 1-23 (1977).

15. F. Wilczek, Magnetic-Flux, Angular-Momentum, and Statistics. *Physical Review Letters* **48**, 1144-1146 (1982).
16. B. I. Halperin, Statistics of Quasiparticles and the Hierarchy of Fractional Quantized Hall States. *Physical Review Letters* **52**, 1583-1586 (1984).
17. D. Arovas, J. R. Schrieffer, F. Wilczek, Fractional Statistics and The Quantum Hall-Effect. *Physical Review Letters* **53**, 722-723 (1984).
18. R. B. Laughlin, Anomalous quantum Hall effect - an incompressible quantum fluid with fractionally charged excitations. *Physical Review Letters* **50**, 1395-1398 (1983).
19. R. Schuster *et al.*, Phase measurement in a quantum dot via a double-slit interference experiment. *Nature* **385**, 417-420 (1997).
20. Y. M. Zhang *et al.*, Distinct signatures for Coulomb blockade and Aharonov-Bohm interference in electronic Fabry-Perot interferometers. *Physical Review B* **79**, 241304 (2009).
21. C. D. C. Chamon, D. E. Freed, S. A. Kivelson, S. L. Sondhi, X. G. Wen, Two point-contact interferometer for quantum Hall systems. *Physical Review B* **55**, 2331-2343 (1997).
22. D. T. McClure, W. Chang, C. M. Marcus, L. N. Pfeiffer, K. W. West, Fabry-Perot interferometry with fractional charges. *Physical Review Letters* **108**, 256804 (2012).
23. H. K. Choi *et al.*, Robust electron pairing in the integer quantum hall effect regime. *Nature Communications* **6**, 7435 (2015).
24. S. L. J. Nakamura, G. C. Gardner & M. J. Manfra, Direct observation of anyonic braiding statistics. *Nature Physics* **16**, 931–936 (2020).
25. Y. Ji *et al.*, An electronic Mach-Zehnder interferometer. *Nature* **422**, 415-418 (2003).
26. I. Neder, M. Heiblum, Y. Levinson, D. Mahalu, V. Umansky, Unexpected behavior in a two-path electron interferometer. *Physical Review Letters* **96**, 016804 (2006).
27. I. Neder *et al.*, Interference between two indistinguishable electrons from independent sources. *Nature* **448**, 333-337 (2007).
28. I. Neder, M. Heiblum, D. Mahalu, V. Umansky, Entanglement, dephasing, and phase recovery via cross-correlation measurements of electrons. *Physical Review Letters* **98**, 036803 (2007).
29. P. Roulleau *et al.*, Finite bias visibility of the electronic Mach-Zehnder interferometer. *Physical Review B* **76**, 161309 (2007).

30. K. T. Law, D. E. Feldman, Y. Gefen, Electronic Mach-Zehnder interferometer as a tool to probe fractional statistics. *Physical Review B* **74**, 045319 (2006).

31. D. E. Feldman, A. Kitaev, Detecting non-Abelian statistics with an electronic Mach-Zehnder interferometer. *Physical Review Letters* **97**, 186803 (2006).

32. B. I. Halperin, A. Stern, I. Neder, B. Rosenow, Theory of the Fabry-Perot quantum Hall interferometer. *Physical Review B* **83**, 155440 (2011).

33. B. Rosenow, B. I. Halperin, Influence of interactions on flux and back-gate period of quantum Hall interferometers. *Physical Review Letters* **98**, 106801 (2007).

34. P. Roulleau *et al.*, Direct measurement of the coherence length of edge states in the integer quantum Hall regime. *Physical Review Letters* **100**, 126802 (2008).

35. R. Bhattacharyya, M. Banerjee, M. Heiblum, D. Mahalu, V. Umansky, Melting of interference in the fractional quantum Hall effect: Appearance of neutral modes. *Physical Review Letters* **122**, 246801 (2019).

36. I. Gurman, R. Sabo, M. Heiblum, V. Umansky, D. Mahalu, Dephasing of an electronic two-path interferometer. *Physical Review B* **93**, 121412 (2016).

37. M. Goldstein, Y. Gefen, Suppression of Interference in Quantum Hall Mach-Zehnder Geometry by Upstream Neutral Modes. *Physical Review Letters* **117**, 276804 (2016).

38. H. Inoue *et al.*, Proliferation of neutral modes in fractional quantum Hall states. *Nature Communications* **5**, 4067 (2014).

39. S. Biswas *et al.*, Does shot noise always provide the quasiparticle charge? *arXiv:2111.05575*, (2021).

40. U. Khanna, M. Goldstein, Y. Gefen, Emergence of Neutral Modes in Laughlin-like Fractional Quantum Hall Phases. *arXiv:2109.15293*, (2021).

41. U. Khanna, M. Goldstein, Y. Gefen, Fractional edge reconstruction in integer quantum Hall phases. *Physical Review B* **103**, L121302 (2021).

42. B. Rosenow, A. Stern, Flux Superperiods and Periodicity Transitions in Quantum Hall Interferometers. *Physical Review Letters* **124**, 106805 (2020).

43. V. V. Ponomarenko, D. V. Averin, Mach-Zehnder interferometer in the fractional quantum Hall regime. *Physical Review Letters* **99**, 066803 (2007).

44. D. E. Feldman, Y. Gefen, A. Kitaev, K. T. Law, A. Stern, Shot noise in an anyonic Mach-Zehnder interferometer. *Physical Review B* **76**, 085333 (2007).
10

45. C. L. Kane, Telegraph noise and fractional statistics in the quantum Hall effect. *Physical Review Letters* **90**, 226802 (2003).
46. N. Byers, C. N. Yang, Theoretical considerations concerning quantized magnetic flux in superconducting cylinders. *Physical Review Letters* **7**, 46 (1961).

**Acknowledgments:** We acknowledge Mitali Banerjee, Ankur Das, Dima E. Feldman, Yuval Gefen, Izhar Neder and Ady Stern for useful discussions and continuous support of the Sub-Micron Center staff. MH acknowledges the support of the European Research Council under the European Community's Seventh Framework Program (FP7/2007-2013)/ERC under grant agreement number 713351 and the partial support of the Minerva foundation under grant number 713534.

**Author contribution:** HKK and SB designed and fabricated the devices. HKK and SB performed the measurements and analysed the data with NO. MH supervised the experiment and analysis. VU developed and grew the heterostructure supporting the 2DEG. All authors contributed to write the manuscript.

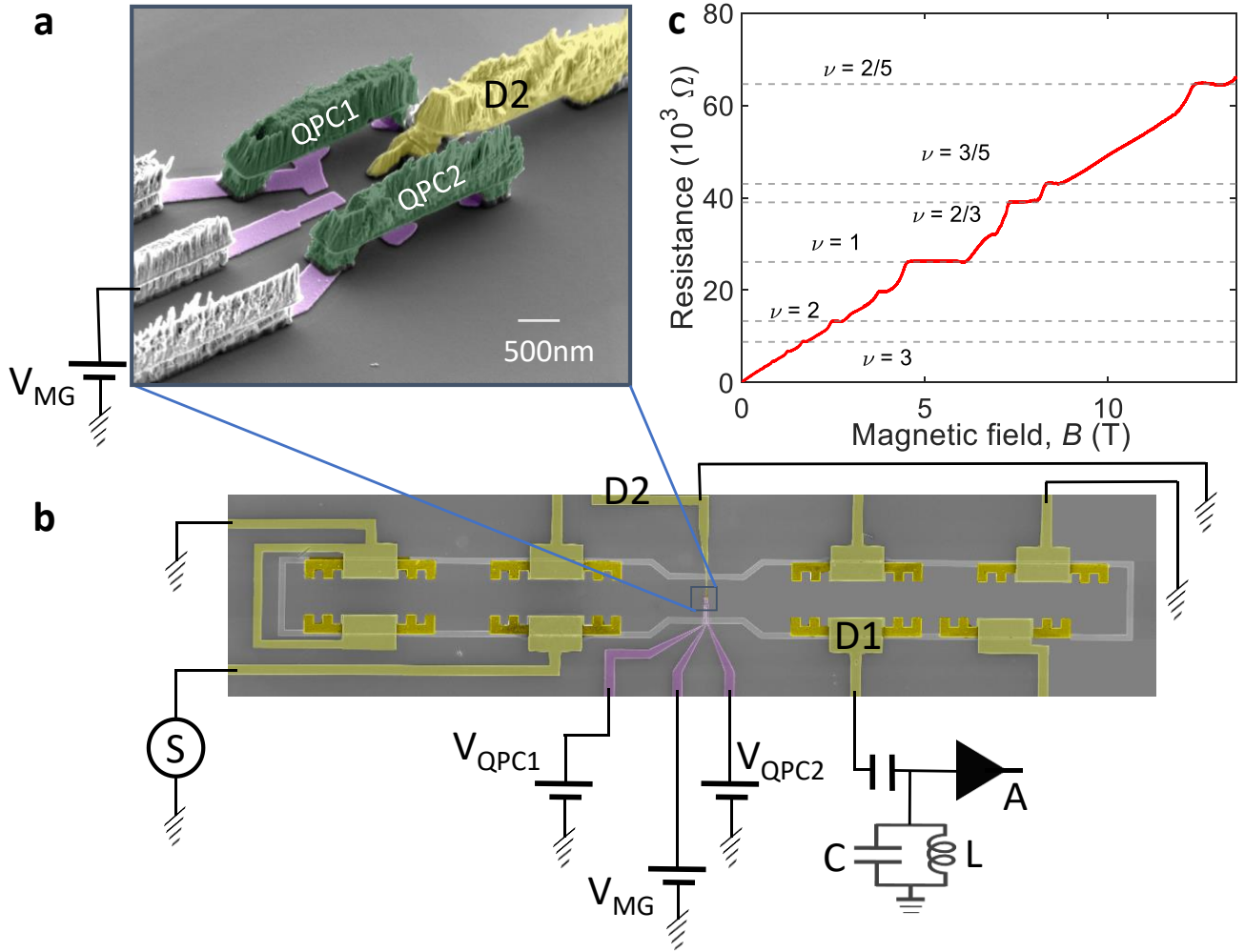

**Fig. 1: Device structure and conductance quantization.** (a) SEM image of the small MZI with 'two-path' area ~3.67μm². Air bridges (green) short the split-gate of the QPCs. Air bridge (yellow) connects drain D2 to ground. The modulation gate repels charge and thus changes the threaded flux in the MZI. (b) Optical image of the full device structure. The MZI shown in (a) is located in the small box in the center. We measured the transmission of MZI at drain D1. (c) Two-probe Hall resistance as a function of the magnetic field. Distinct quantization of quantum Hall plateaus are observed.

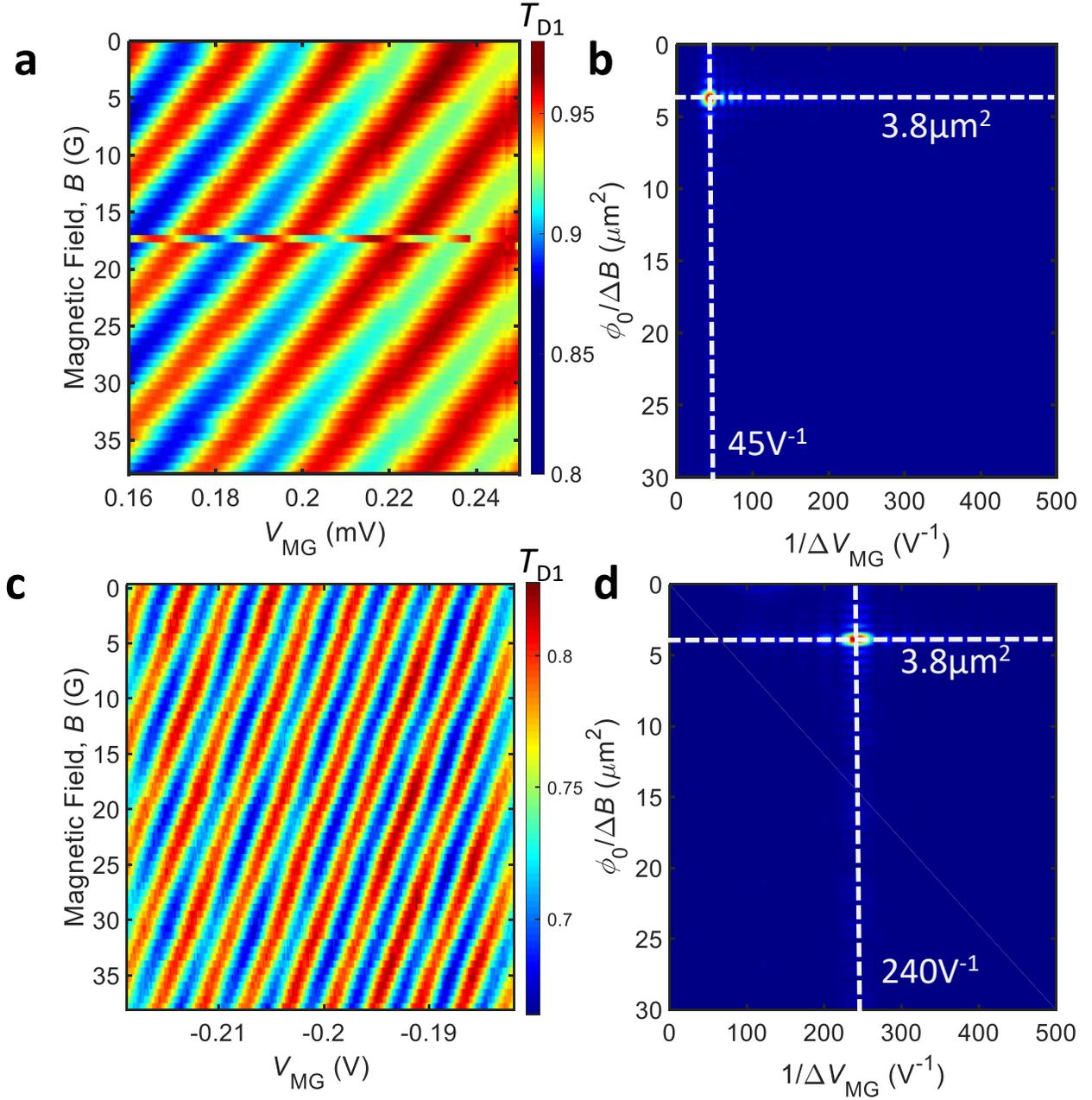

**Fig. 2: Integer and fractional Aharonov-Bohm (AB) interference patterns.** (**a**) & (**c**) '*Pajama* plots' – Transmission of the interfering outer edge modes in the $V_{MG}$-$B$ plane. (**a**) Interference of the outer edge mode in bulk filling $\nu=2$ ($B=2.5$T). (**c**) Interference of the outer edge mode $\nu=1/3$ in bulk filling $\nu=2/5$ ($B=12.65$T). The equiphase lines are typical of AB interference. In both states, the field periodicity is of a single flux quantum. (**b**) & (**d**) 2D Fourier transforms of the 'pajama plots' showing a single peak, excluding Coulomb effects. The two periodicities in $V_{MG}$, 22.2mV in $\nu=2$ and 4.2mV in $\nu=2/5$, are proportional to $1/B$.

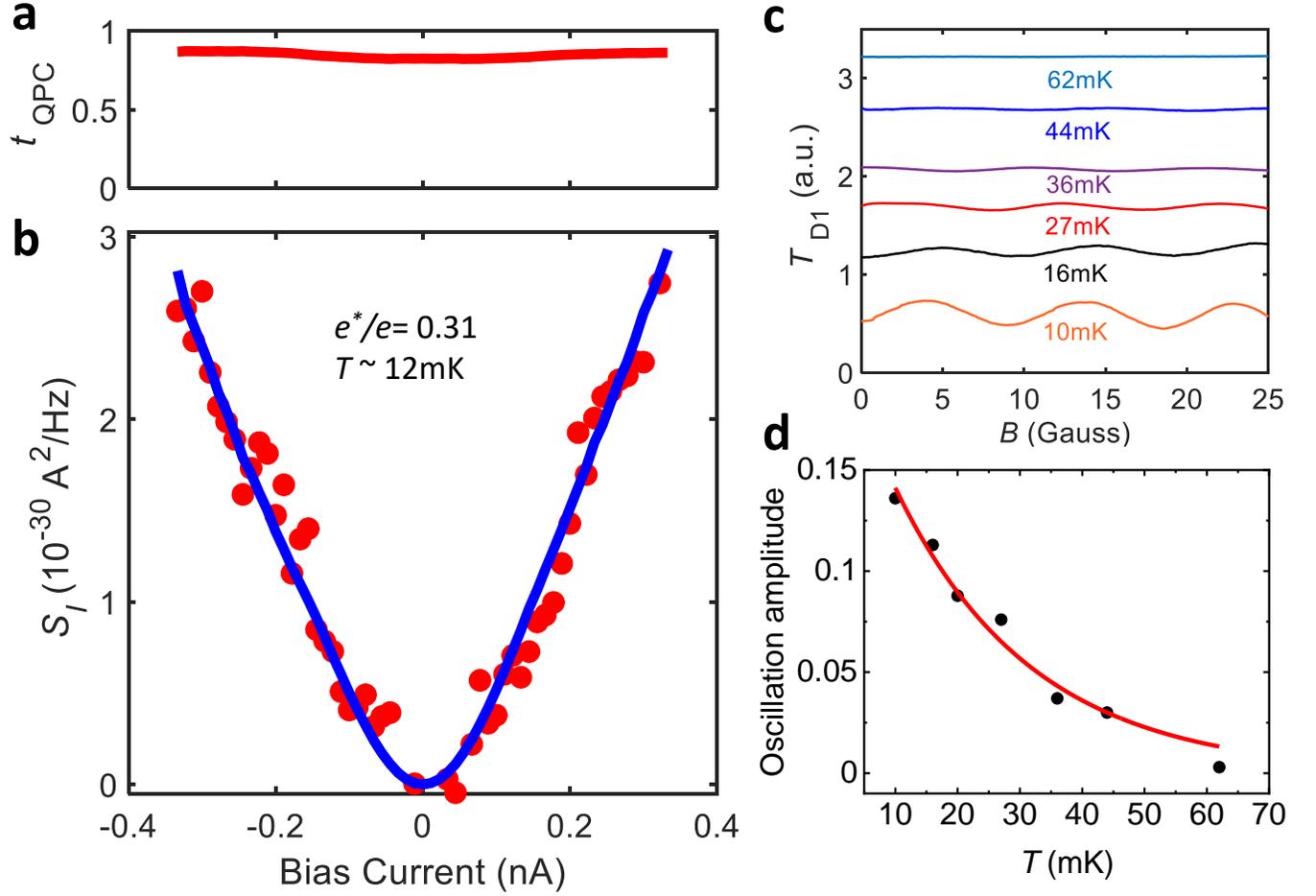

**Fig. 3: Charge determination via shot noise measurements and temperature-dependent visibility:** **(a)** Non-linear differential transmission of the partitioned 1/3 outer edge mode by a QPC (at bulk filling $\nu$=2/5), where $t_{QPC}$~0.82. **(b)** Current dependent spectral density of the shot noise (red discrete data points). The fit (blue line) agrees with charge $e^*$~0.31$e$ at electron temperature ~12mK. **(c)** Temperature dependence of the $B$-dependent interfering traces of the 1/3 mode. **(d)** An exponential fit of the temperature-dependent interfering oscillations leads to a characteristic temperature of ~23mK.

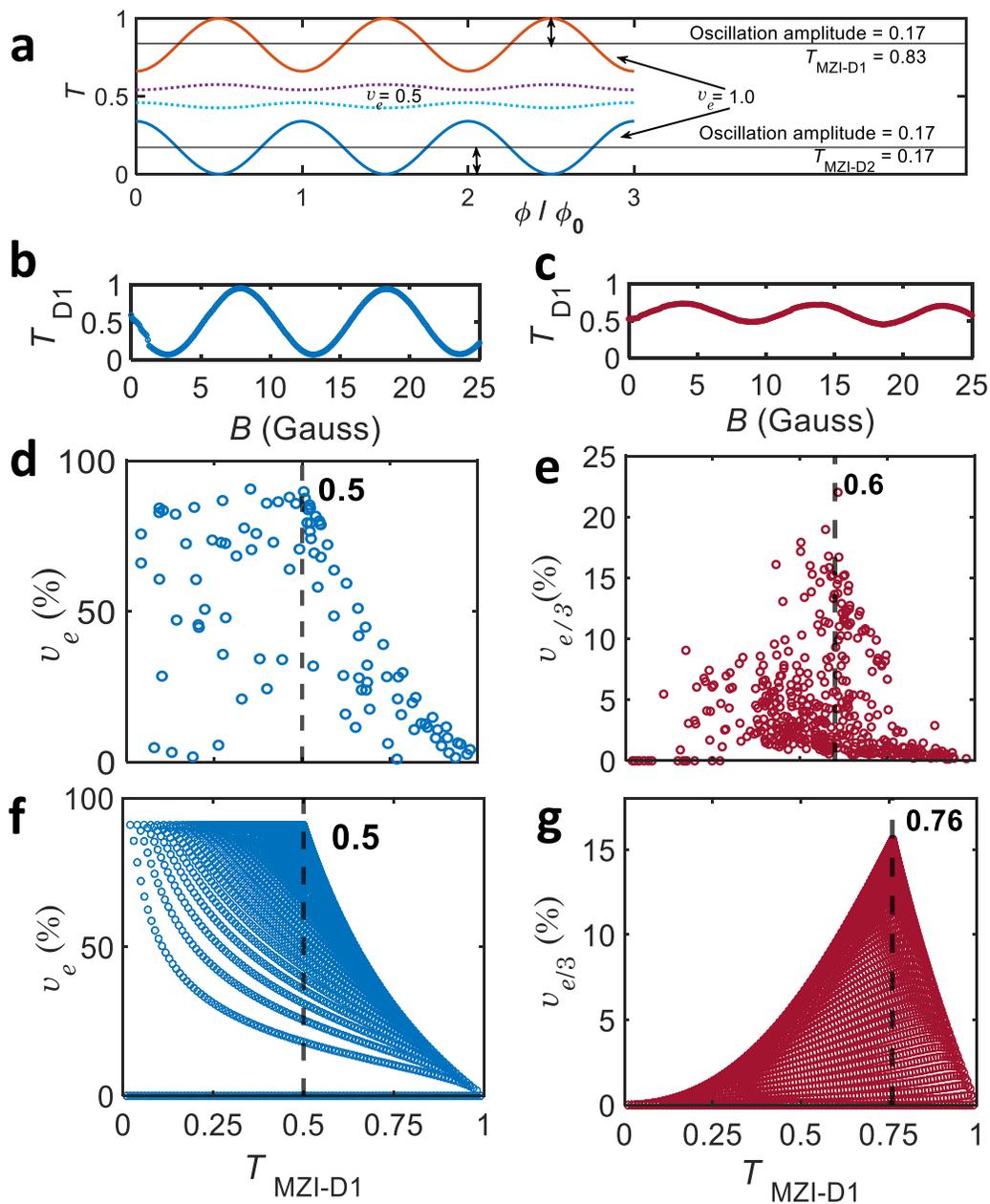

**Fig. 4: Visibility in the integer and fractional regimes. (a)** Calculated anyonic interference patterns in the two drains, D1 and D2 (Eqs. (1) & (2)). Solid orange and blue lines represent the maximum interference amplitude (at $|t_i|^2=0.5$), which is expected at D1 and at D2 for the case of an ideal visibility in the integer regime, is $v_e = 1$. Though the oscillation amplitudes at D1 and D2 (but out of phase) are same, the average transmissions are very different ($T_{MZI-D1} \sim 5 \times T_{MZI-D2}$, two solid black lines). This forces the visibility at D1 (20%, $T_{MZI} \sim 0.83$) to be ~5 times smaller than at D2 (100%, $T_{MZI} \sim 0.17$). The visibility of the interfering 1/3 outer mode drops off rapidly with diminishing integer visibility, $v_e$. An example is shown by the dotted oscillations (in purple and cyan) for $v_e = 0.5$. **(b)** Traces of measured interference oscillation of the interfering outer $v=1$ mode at bulk filling $v=2$ ($B=2.5$T). **(c)** A similar plot of the measured interfering outer $v=1/3$ mode at bulk filling $v=2/5$ ($B=12.65$T). **(d & e)** The measured visibilities in the integer and fractional regimes as a function the average transmission $T_{MZI-D1}$, respectively. Each dot represents a different transmission combination, ($|t_1|^2$, $|t_2|^2$), of the two QPCs. In the fractional regime, the QPCs' individual transmissions were kept relatively high around the peak value of the visibility to assure a Fano factor=1/3 at each QPC. **(f & g)** Calculated visibility at D1 for integer and fractional regimes, respectively. The points in a line shape are for single $|t_1|^2$, while $|t_2|^2$ is varied from 0 to 1 in steps. Different lines are corresponding to multiple values of $|t_1|^2$, covering both QPCs limits. Note that each point has appeared twice for a ($|t_1|^2$, $|t_2|^2$) and in reverse. The striking difference between these two visibilities emanates from the different braiding phases of the electrons and the fractional charges (see also details section SM 7). The measured $v_e$ dependence with $T_{MZI-D1}$ is in accordance with $\eta=0.91$ (d & f). Considering same $\eta$, $v_{e/3}$ profile with $T_{MZI-D1}$ matches fairly well, though there is a slight disagreement in details of peak $v_{e/3}$. We find the peak $v_{e/3}$ (~22%) at $T_{MZI-D1} \sim 0.6$, where as the expected peak is at 0.76 with $v_{e/3} = 15.6\%$.

# Supplementary Material

# Anyonic interference and braiding phase in a Mach-Zehnder Interferometer

**SM1: Fabrication details:** The Hall bars were patterned by wet etching of the GaAs/AlGaAs heterostructure. The 2DEG depth below the surface was 103nm, spacer layer (separation between donors and 2DEG) was 70nm and the quantum well width was 30nm. Ohmic contacts were made by alloying the following sequence of evaporation (from GaAs surface and up): Ni(5nm)/Au(220nm)/Ge(110nm)/Ni(83nm)/Au(20nm). The gates used to pattern the MZI were obtained by evaporating Pd-Au(10nm)/Au(10nm) in ultra-high vacuum. During the gate evaporation the sample is kept at liquid $N_2$ temperature. In the last step air bridges were made by evaporating Ti(25nm)/Au(480nm) to connect the inner drain, D2 and the gates forming the QPCs. SEM image of the device with interfering edges are shown in Fig. SM1. We bias cooled the gates with +0.5V.

**SM2: Interference at $\nu=3$:** The measured interference of the outer edge mode at bulk filling $\nu=3$. The observed interference pattern in $B$-$V_{MG}$ (Pajama) plane is plotted in Figs. SM2 (a) & (b) for small and large MZI, respectively. The flux periodicity is $\sim\phi_0$ as detailed in the **Table** (main text).

**SM3: Modulation gate dependence:** The purpose of tuning the modulation gate is to change the area enclosed by the interferometer and hence the enclosed flux. The charge expelled by biasing the modulation gate,

$$\Delta q = C\Delta V_{MG} = n_e e \Delta A \ ,$$

where $\Delta A$ is the change in enclosed in MZI area, and $n_e$ the electrons density. Change in the area $\Delta A$ leads to change in the AB phase of interference,

$$\Delta\theta = 2\pi \frac{e^*}{e} \frac{B.\Delta A}{\phi_0} \ .$$

Over a full period of interference the phase change is $2\pi$, hence combining this two equations,

$$\frac{e^*}{e} = \frac{1}{B \cdot \Delta V_{MG}} \frac{\phi_0 n_e e}{C} .$$

While we compare interference of two bulk filling factors with interfering charge $e_1^*$ and $e_2^*$ and modulation gate periodicity $\Delta V_{MG1}$ and $\Delta V_{MG2}$, respectively at magnetic field $B_1$ and $B_2$, which changes the flux by $\Delta \phi_1$ and $\Delta \phi_2$ the relation obtains,

$$\frac{\Delta \phi_2}{\Delta \phi_1} = \frac{e_1^*}{e_2^*} = \frac{B_2 \Delta V_{MG2}}{B_1 \Delta V_{MG1}} .$$

This relation holds with assumption that the capacitance between modulation gate and edge channel remain constant with *B*.

**SM4: Interference as we approach *v*=1:** Even though high visibility interference is observed at outer modes, we are unable to observe interference at innermost modes. We also do not observe interference at *v*=1, where the interfering mode faces the bulk. Increasing field from *v*=4/3 state (*B*=3.675T) towards *v*=1 (*B*=5.0T), the visibility diminishes from ~17% to zero. Fig. SM3 shows characteristic interference traces with *B*, while moving from *v*=4/3 state to *v*=1 state (intermediate point *B*=4.3T). This phenomenon was observed before and assumed to be a result of the emerging neutral modes due to edge reconstruction (*1, 2*).

**SM5: Edge modes at *v*=2/5:** The 2/5 FQHE state supports two edge modes with conductance $e^2/15h$ of the inner mode and $e^2/3h$ of the outer mode. These two modes can be identified in the QPC response. This is shown in Fig. SM4.

**SM6: Noise measurement at *v*=2/5:** Shot noise in a single QPC was measured to obtain the quasiparticle charge. The spectral density of shot noise is, $S_I = 2FeI_{dc}t_{QPC}(1 - t_{QPC})[\coth\left(\frac{FeV}{2k_BT}\right) - \frac{2k_BT}{FeV}]$, where *F* is Fano factor (in most cases is the quasiparticle charge, $t_{QPC}$ is the transmission of the QPC, and $I_{dc}$ is the impinging DC current) (*3-7*). The charge was 0.31*e* in the partitioned 1/3 mode. Partitioning the inner 1/15 edge led to the shot noise plotted in Fig. SM5 (red dots) lower panel, while the non-linear transmission dependence on the bias is plotted in upper panel. The blue solid line is the fit to the measured noise. The extracted partitioning charge is 0.22$e$ at electron temperature ~12mK. We did not find excess noise on the QPC-plateau of 1/3 in 2/5 state at base temperature, which shows an absence of neutral mode (*8*). This is consistent with the significant visibility in interference.

**SM7: Theoretical understanding of fractional interference in MZI:** We outline in some detail the model used for quasiparticles (QPs) charge *e*/m. The assumption in the model is that *e*/m quasiparticles can only get absorbed in drain after bunching to an electron (*9, 10*). The first (m-1) QP will be stuck before m$^{th}$ QP arrives. This has a significant effect in fractional interference of MZI. Presence of any QP at the D2 drain (inside the MZI), changes the probability for the next QP's acquired phase. The interfering QP encircles D$_2$ and the QP stuck at D2 gives rise to an extra exchange phase on top of the bare AB phase. This process compels the transmission of individual QP in MZI to be a correlated phenomenon (modulo m). Measured interference of QP in MZI is hence an averaged probability of m correlated QP transmission compared to individual independent electron transmission (IQH).

The probability of an 1/m QP to reach D2 can generally be written as $\bar{p}\left(1 + v_e \cdot \cos\left(\frac{2\pi\phi}{m\phi_0} + \frac{2\pi}{m} \cdot n\right)\right)$, where $\bar{p}$ is average probability and $n = 0, 1, \ldots, (m-1)$. For electron, m=1 ($n = 0$), it obtains the discussed probability or transmission of MZI. In case of 1/3 QP (m=3), the same probability is $p_n = \bar{p}\left(1 + v_e \cdot \cos\left(\frac{2\pi\phi}{3\phi_0} + \frac{2\pi}{3} \cdot n\right)\right)$ for $(n+1)^{th}$ QP to reach D2, where $n = 0, 1, 2$. The $\frac{2\pi}{3}n$ factor is because while $n$ number of QP are stuck at inner drain and the next interfering QP encircles the $n$ QP at D2 (exchange phase). This contributes to a $\frac{2\pi}{3} \cdot n$ phase shift in probability of next QP to reach D2. The time requires to arrive $(n+1)^{th}$ QP to reach D2 is $t_n = 1/p_n$. In this process when there are 3 QP arrives at D2, it gets absorbed in drain restoring the MZI to initial condition. Here $p_n$ is not the transmission we can measure at D2 because the drain can only absorb once 3 of these 1/3 QPs accumulates and become an electron(*9-12*).

The measured transmission at D2 would be the harmonic average of this single 1/m QP probabilities,

$$T_{D2} \equiv m. \left[\sum_{n=0}^{m-1} 1/p_n\right]^{-1}$$

$$= m\left[\sum_{n=0}^{m-1} \frac{1}{\bar{p}\,(1+v_e.\cos(a_n))}\right]^{-1}$$

where $a_n = \frac{2\pi\phi}{m\phi_0} + \frac{2\pi n}{m}$

$$= m\left[\sum_{n=0}^{m-1} \frac{1}{\bar{p}\,(1+v_e.\cos(a_n))} \cdot \frac{M+\cos(m.a_n)}{M+\cos(m.a_n)}\right]^{-1}$$

The polynomial of the form $M + \cos(m.a_n)$, can be simplified to $M + \cos\left(\frac{2\pi\phi}{\phi_0}\right)$. The polynomial is chosen such a way that it exactly cancels with $(1 + v.\cos(a_n))$.

$$= m\left[\sum_{n=0}^{m-1} \frac{1}{\bar{p}\,\left(M+\cos\left(\frac{2\pi\phi}{\phi_0}\right)\right)} \cdot \frac{M+\cos(m.a_n)}{(1+v_e.\cos(a_n))}\right]^{-1}$$

$$= m\left[\frac{1}{p_0\left(M+\cos\left(\frac{2\pi\phi}{\phi_0}\right)\right)} \sum_{n=0}^{m-1} \frac{\sum_{s=0}^{m} A_s \cos^s a_n}{(1+v_e.\cos(a_n))}\right]^{-1}$$

$$= m\left[\frac{1}{\bar{p}\,\left(M+\cos\left(\frac{2\pi\phi}{\phi_0}\right)\right)} \sum_{n=0}^{m-1} \frac{\sum_{t=0}^{m-1} B_t \cos^t(a_n).(1+v.\cos(a_n))}{(1+v_e.\cos(a_n))}\right]^{-1}$$

$$= m\left[\frac{1}{\bar{p}\,\left(M+\cos\left(\frac{2\pi\phi}{\phi_0}\right)\right)} \sum_{n=0}^{m-1}\sum_{t=0}^{m-1} B_t \cos^t a_n\right]^{-1}$$

$$= m\left[\frac{1}{\bar{p}\,\left(M+\cos\left(\frac{2\pi\phi}{\phi_0}\right)\right)} \sum_{u=0}^{m-1}\sum_{n=0}^{m-1} C_u \cos(u.a_n)\right]^{-1}$$

Using the relation of summing nth roots of unity to zero, only n=0 contributes.

$$T_{D2} = \frac{\bar{p}}{C_0}\left(M + \cos\left(\frac{2\pi\phi}{\phi_0}\right)\right)$$

In case of 1/3 QP, m=3. Then the polynomial $M + \cos\left(\frac{2\pi\phi}{\phi_0}\right)$ takes the form for $M = \frac{4-3v_e^2}{v_e^3}$ which obtains $C_0 = \frac{4-v_e^2}{v_e^3}$. Putting these values,

$$T_{D2} = (|t_1 r_2|^2 + |r_1 t_2|^2)\frac{4 - 3v_e^2}{4 - v_e^2}\left[1 + \frac{v_e^3}{4 - 3v_e^2}\cos\left(\frac{2\pi\phi}{\phi_0}\right)\right]$$

**SM8: Fractional interference for the larger MZI (area ~13.5μm²).** Interference in this MZI in the integer regime shows a visibility $v_e$~67% at $v=2$ outer edge. The interference is plotted in Fig. SM6. Likewise, interference at outer 1/3 edge mode with the bulk filling $v=2/5$ plotted in Fig. SM7. The maximum visibility ($v_{e/3}$) was ~8.3%. The variation of visibility in case of fractional interference with average transmission of MZI ($T_{MZI}$) is plotted in Fig. SM7. As shown in section SM8, the peak visibility appears at $T_{MZI}$=0.54.

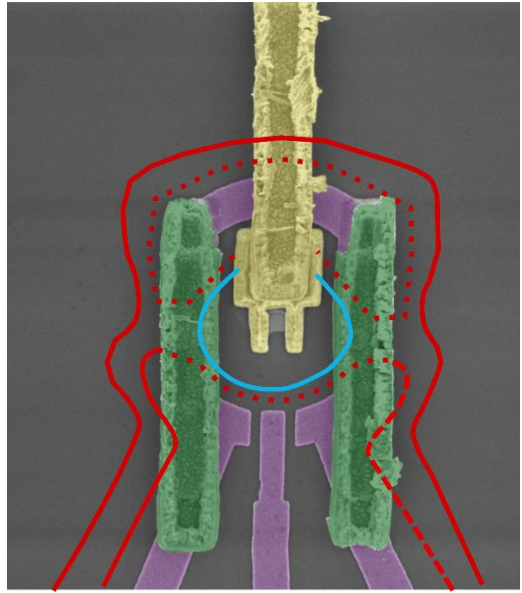

**Fig. SM1: Device Image.** SEM image of the small MZI with 'two-path' area ~3.67μm$^2$. The paths of the two edge modes in ν=2 (similarly in ν=2/5) are drawn, with the inner mode is fully reflected by the two QPCs and outer mode interferes (shown by dotted red lines). Full lines, blue (cold) and red (hot), are unpartitioned modes.

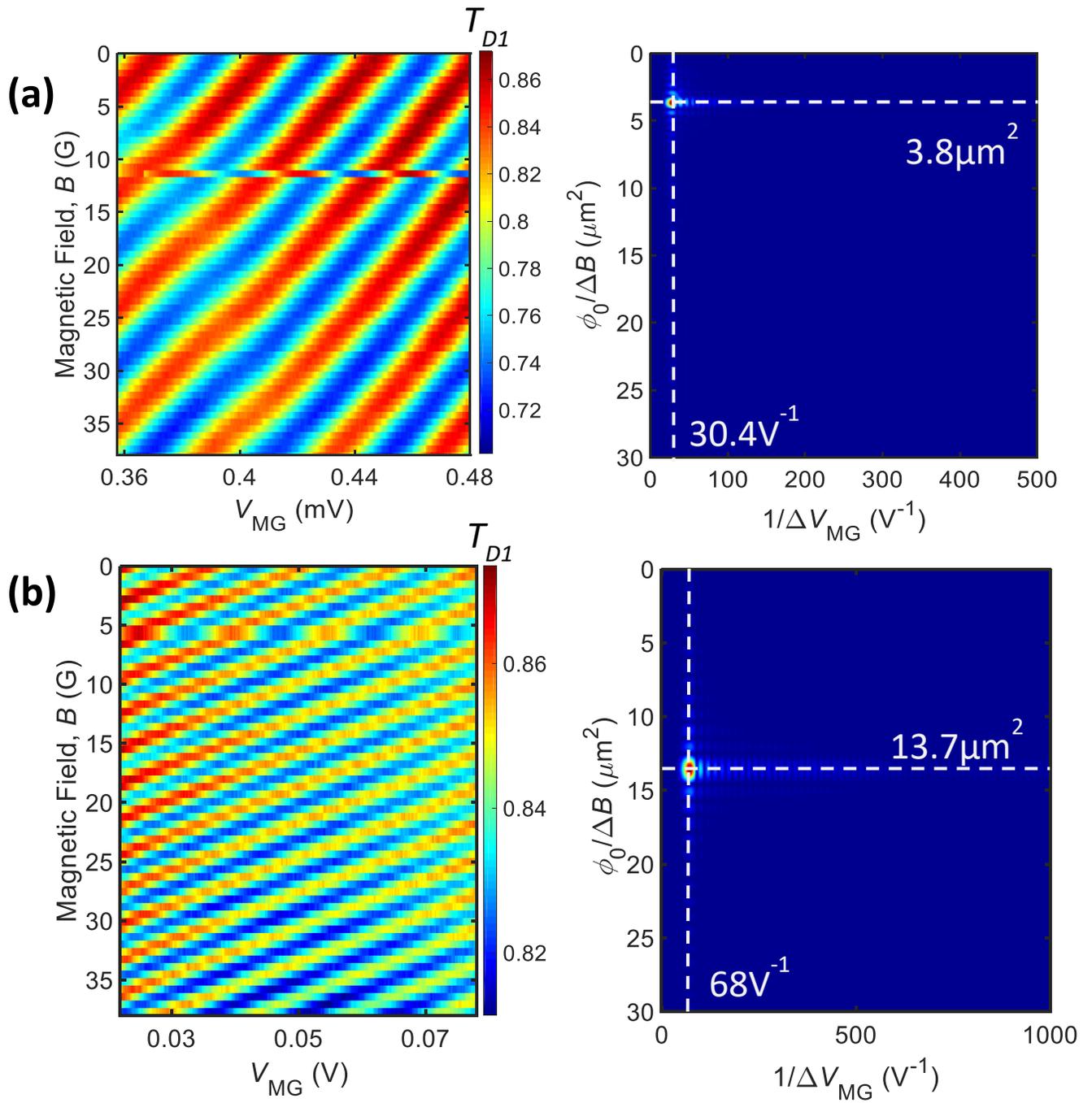

**Fig. SM2: Interference at outer edge of *v*=3:** (a) & (b) 'Pajama plot' (V$_{MG}$ – B 2D plane) shows interference outer edge of *v*=3 for small and large MZI. Right panel shows the 2D- Fourier transform of the interference, showing one peak corresponding to the periodicity of 2-path interference. The flux periodicity corresponds to one flux quanta.

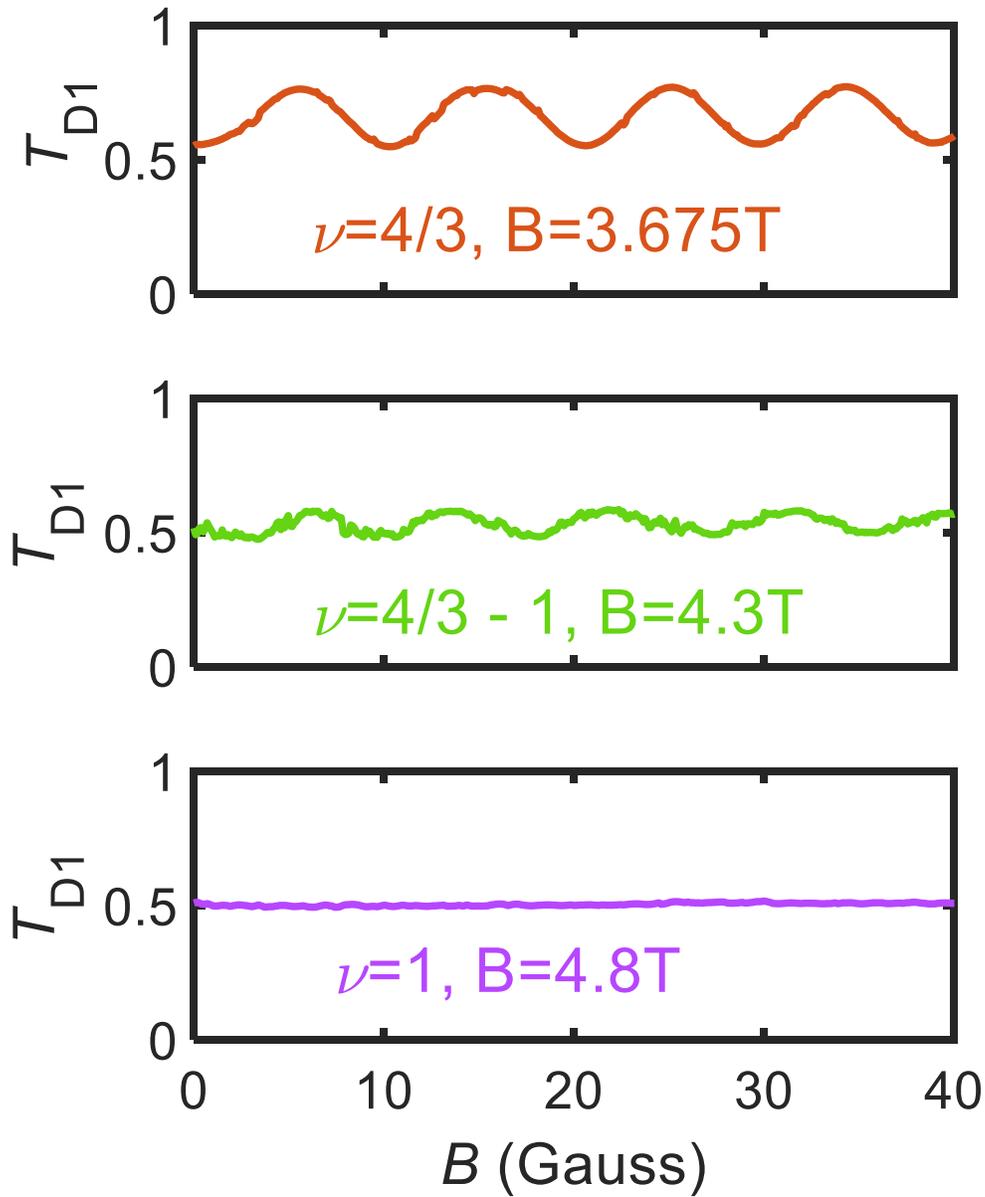

**Fig. SM3: Interference close to ν=1:** Traces of interference as we change the flux enclosed by changing *B*. (a) shows measured interference at *v*=4/3 (*B*=3.675T) outer edge (*v*=1) with visibility ~17%. As we increase *B*, at *B*=4.3T visibility degrades and eventually goes to zero at *v*=1. In the entire plateau of *v*=1, from *B*=4.8T to 5.3T, we did not find interference.

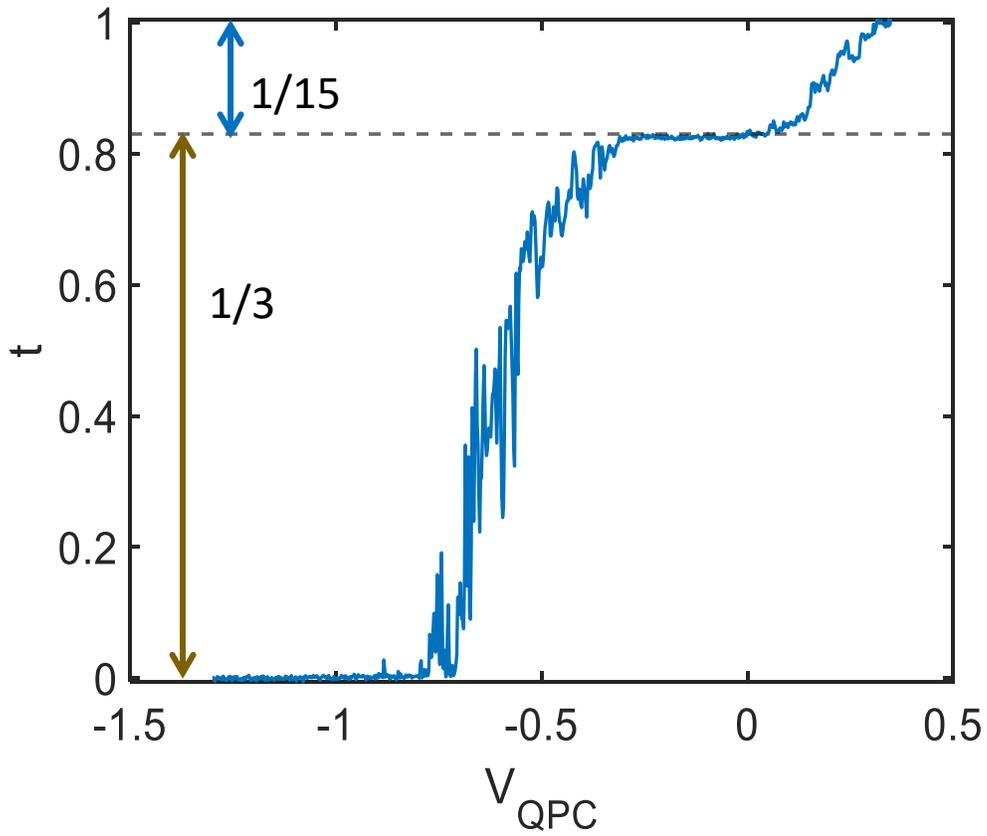

**Fig. SM4: QPC response of *v*=2/5:** QPC transmission of 2/5 edge mode as a function of QPC gate voltage ($V_{QPC}$). The $(2/5)e^2/h$ full transmission (t=1) followed by $(1/3)e^2/h$ intermediate plateau while the inner $(1/15)\ e^2/h$ channel is fully reflected (t=0.83) as we pinch the QPC.

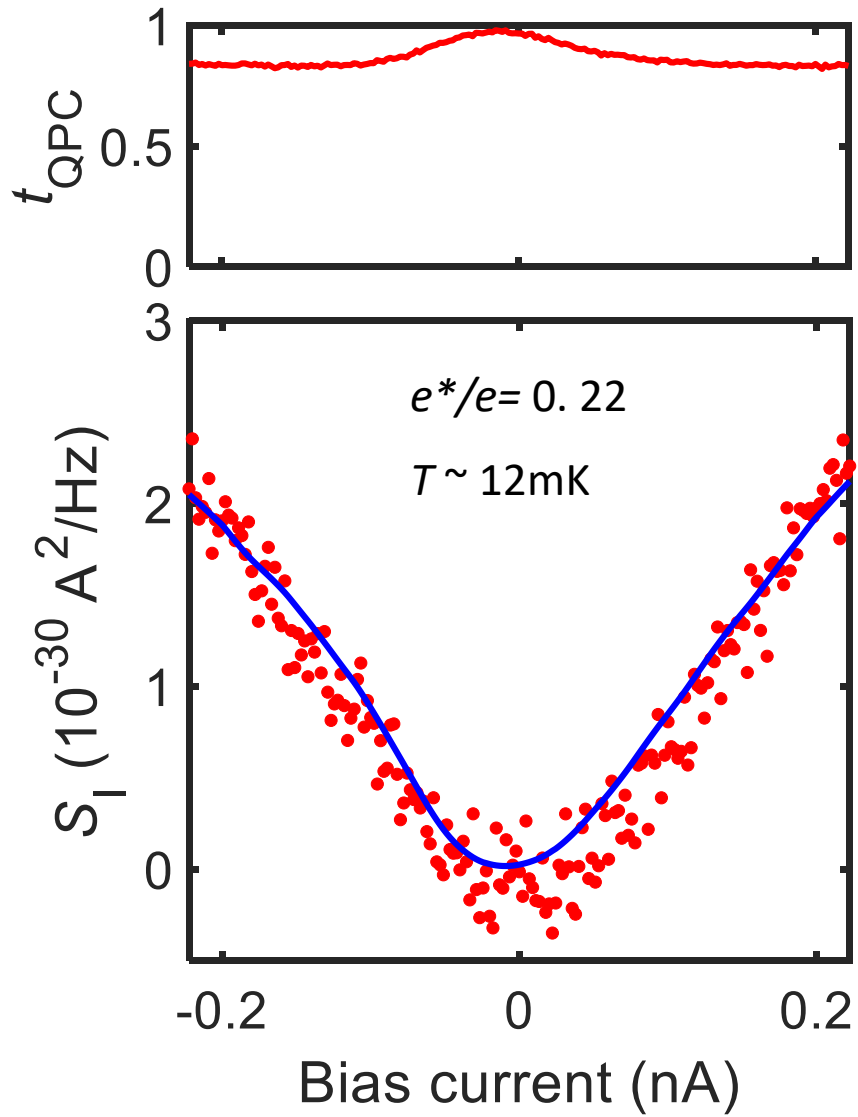

**Fig. SM5: Shot-noise measurement at 1/15 edge channel:** Measured non-linear bias dependence (red plot, upper panel) and shot noise data (red dots, lower panel) while partitioning inner 1/15 mode. The weak back scattering in QPC was set ~4%. The blue solid line is the fit to the noise data which obtains fano-factor to be ~0.22 and temperature ~12mK.

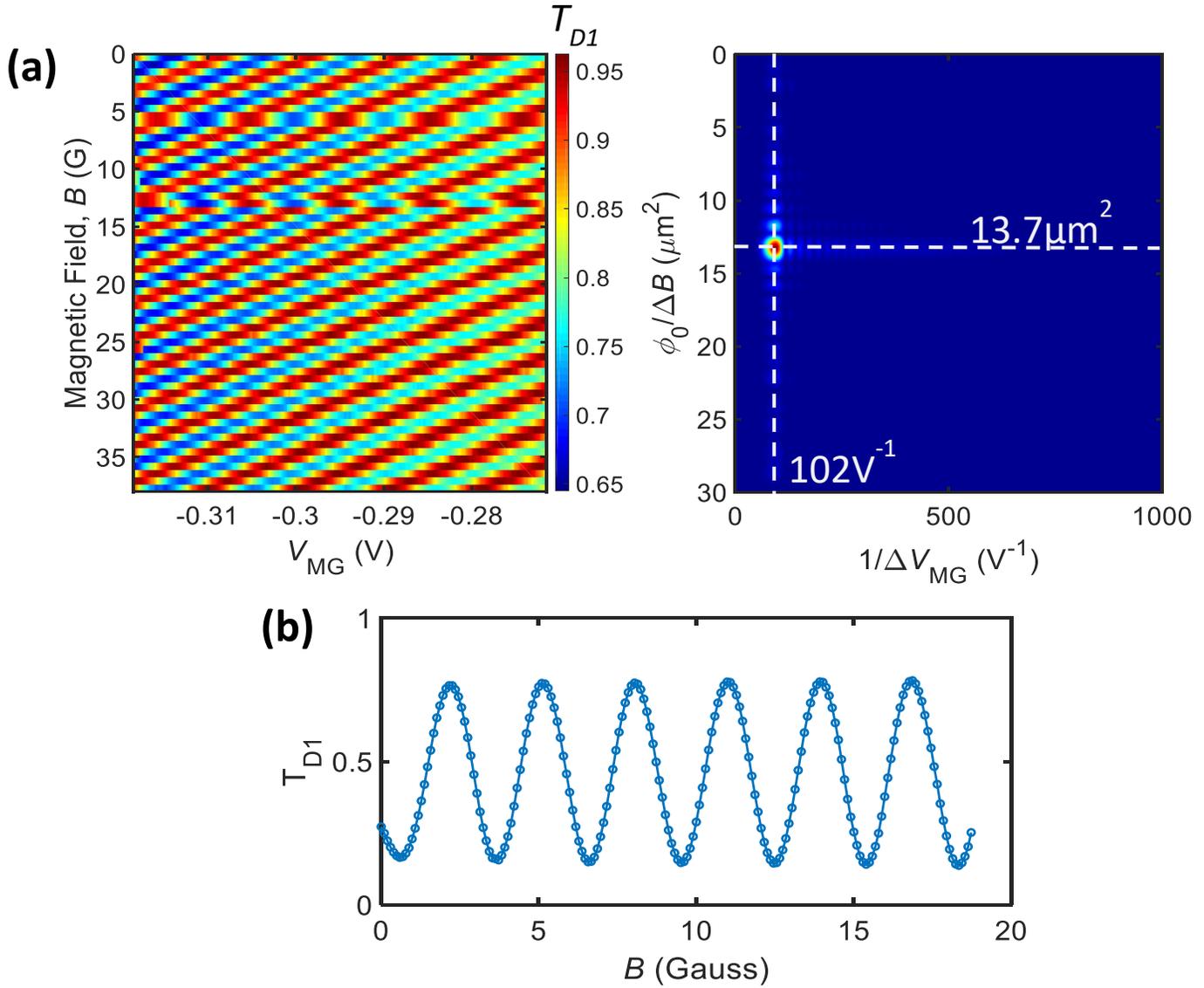

**Fig. SM6: Interference in large MZI:** (a) 2D plot of interference of large (13.5μm$^2$) MZI as a function of modulation gate and magnetic field at bulk filling $v=2$ while interfering outer edge. Right panel shows Fourier transformation of interference measured showing one peak in periodicity. The AB type interference showed same periodicity ($\Delta B$ = 3.2 Gauss) in integer states obtaining flux periodicity of $\phi_0$. (b) Line trace of interference oscillation of outer edge at $v=2$ with 67% visibility as the magnetic field is tuned.

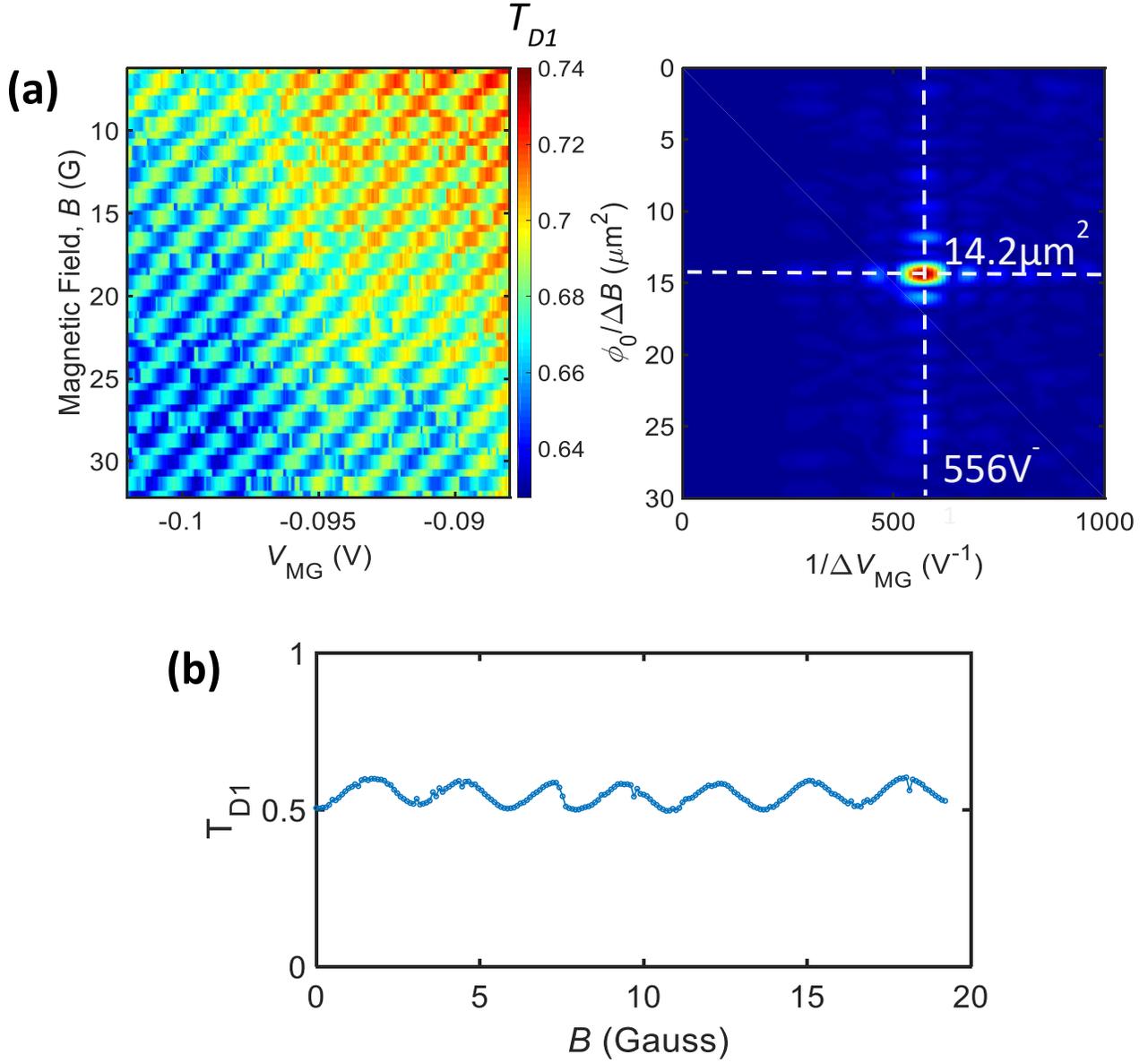

**Fig. SM7: Interference in large MZI:** (a) 2D plot of transmission of MZI as a function of modulation gate and magnetic field at bulk filling $v=2/5$ while interfering outer 1/3 edge. Fourier transformation of interference measured showing one peak in periodicity, shown in right. The AB type interference showed same periodicity ($\Delta B = 2.9$ Gauss) as in integer states obtaining flux periodicity of $\phi_0$. (b) Trace of interference oscillation measured for interfering 1/3 edge with B.

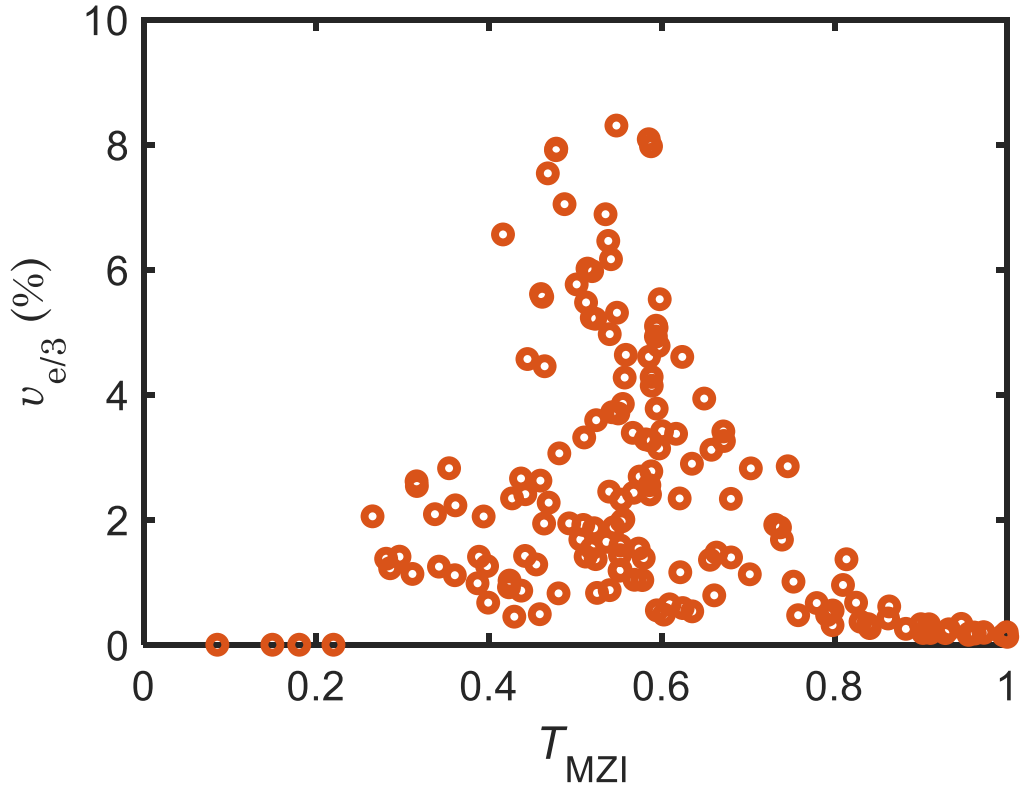

**Fig. SM8: Visibility profile with $T_{MZI}$ for larger MZI:** Measured visibility at D1 in fractional interference of 1/3 edge at bulk filling of $v=2/5$. Similar to what we expect, we observe a peak like behavior in visibility with average transmission, $T_{MZI}$. The peak visibility, ($v_{e/3}$), is ~8.3% at $T_{MZI}$ ~0.54. The lower visibility than the smaller MZI and the peak closer to $T_{MZI}$ ~0.5 agrees with our interpretation of fractional interference in MZI.

**References for Supplementary Material:**

1. R. Bhattacharyya, M. Banerjee, M. Heiblum, D. Mahalu, V. Umansky, Melting of interference in the fractional quantum Hall effect: Appearance of neutral modes. *Physical Review Letters* **122**, 246801 (2019).
2. I. Gurman, R. Sabo, M. Heiblum, V. Umansky, D. Mahalu, Dephasing of an electronic two-path interferometer. *Physical Review B* **93**, 121412 (2016).
3. T. Martin, R. Landauer, Wave-packet approach to noise in multichannel mesoscopic systems. *Physical Review B* **45**, 1742-1755 (1992).
4. M. Buttiker, Scattering theory of current and intensity noise correlations in conductors and wave guides. *Physical Review B* **46**, 12485-12507 (1992).
5. R. dePicciotto *et al.*, Direct observation of a fractional charge. *Nature* **389**, 162-164 (1997).
6. A. Bid, N. Ofek, M. Heiblum, V. Umansky, D. Mahalu, Shot noise and charge at the 2/3 composite fractional quantum Hall state. *Physical Review Letters* **103**, 236802 (2009).
7. L. Saminadayar, D. C. Glattli, Y. Jin, B. Etienne, Observation of the e/3 fractionally charged Laughlin quasiparticle. *Physical Review Letters* **79**, 2526-2529 (1997).
8. S. Biswas *et al.*, Does shot noise always provide the quasiparticle charge? *arXiv:2111.05575*, (2021).
9. K. T. Law, D. E. Feldman, Y. Gefen, Electronic Mach-Zehnder interferometer as a tool to probe fractional statistics. *Physical Review B* **74**, 045319 (2006).
10. A. Stern, Anyons and the quantum Hall effect - A pedagogical review. *Annals of Physics* **323**, 204-249 (2008).
11. V. V. Ponomarenko, D. V. Averin, Mach-Zehnder interferometer in the fractional quantum Hall regime. *Physical Review Letters* **99**, 066803 (2007).
12. D. E. Feldman, Y. Gefen, A. Kitaev, K. T. Law, A. Stern, Shot noise in an anyonic Mach-Zehnder interferometer. *Physical Review B* **76**, 085333 (2007).